\documentclass[showpacs,amsmath,amssymb,nofootinbib]{revtex4}
\usepackage{dcolumn}% Align table columns on decimal point
\usepackage{bm}% bold math
\begin{document}

%\preprint{APS/123-QED}

\title{Pairing effect on the giant dipole resonance width at low temperature
\footnote{Accepted in Phys. Rev. C}}

\author{Nguyen Dinh Dang}
\affiliation{%
RI-beam factory project office, RIKEN, 2-1 Hirosawa, Wako, 351-0198 
 Saitama, Japan}%
\author{Akito Arima}
 
\affiliation{House of Councilors, 2-1-1 Nagata-cho, 
Chiyoda-ku, Tokyo 100-8962, Japan}%

\date{\today}% It is always \today, today,
             %  but any date may be explicitly specified

\begin{abstract}
The width of the giant dipole resonance (GDR) at finite temperature $T$
in $^{120}$Sn is calculated within the Phonon Damping Model 
including the neutron thermal pairing 
gap determined from the modified BCS theory. 
It is shown that the effect of thermal pairing causes a smaller GDR 
width at
$T\alt$ 2 MeV as compared to the one obtained neglecting 
pairing. This improves significantly the agreement between theory and 
experiment including the most recent data point at 
$T=$ 1 MeV.
\end{abstract}

% PACS, the Physics and Astronomy
                             % Classification Scheme.
%\keywords{Suggested keywords}%Use showkeys class option if keyword
                              %display desired
\maketitle
\section{Introduction}
Intensive experimental studies of highly-excited nuclei 
during the last two decades have produced 
many data on the evolution of the giant dipole resonance (GDR) 
as a function of temperature $T$ and spin. 
The data show that the GDR width increases sharply 
with increasing $T$ from $T\agt$ 1 MeV up to $\simeq$ 3 MeV. 
At higher $T$ a width saturation 
has been reported (See \cite{Woude} for the most recent review).
The increase of the GDR width with $T$
is described reasonably well within the thermal 
shape-fluctuation model~\cite{TFM} and the phonon damping model 
(PDM)~\cite{PDM1,PDM2,PDM3}. The 
thermal shape-fluctuation model 
assumes an adiabatic coupling of GDR 
to quadrupole degrees of freedom with deformation
parameters $\beta$ and $\gamma$ induced by thermal 
fluctuations and high spins in the intrinsic frame of reference.  
Although this model shows 
an increase of the GDR width with $T$ comparable with the experimental 
systematic at 1.2 $\alt T\alt$ 3 MeV, the GDR shapes generated using 
the strength function of this model differ significantly from 
the experimental ones~\cite{Gervais}.
The PDM considers the coupling of 
the GDR to $pp$ and $hh$ configurations at $T\neq$ 0 as the  
mechanism of the width increase and saturation.  The PDM calculates
the GDR width and strength function directly in the laboratory frame without 
any need for an explicit inclusion of thermal fluctuation of 
shapes.
The PDM reproduces fairly well both of the observed 
width~\cite{PDM1,PDM2} and shape~\cite{PDM3,Kaury} of 
the GDR at $T\neq$ 0.

In general, pairing 
was neglected in the calculations for hot GDR 
as it was believed that the gap vanishes at $T=T_{\rm 
c}<$ 1 MeV according to the temperature BCS theory. However, it 
has been shown in \cite{Moretto} that thermal 
fluctuations smear out the superfluid-normal 
phase transition in finite systems so that the pairing gap survives up to 
$T\gg$ 1 MeV. This has been confirmed microscopically 
in the recent modified 
Hartree-Fock-Bogoliubov (HFB) theory at finite 
$T$~\cite{MHFB}, whose limit is the modified-BCS  
theory~\cite{MBCS,MBCS1}. Other approaches such as the 
static-path approximation~\cite{DangRing}, shell model 
calculations~\cite{Ze}, 
as well as the exact solution of the pairing problem 
\cite{Volya} also show that pairing 
correlations do not abruptly disappear at $T\neq$ 0.
It was suggested in \cite{DGDR} that the decrease of the 
pairing gap with increasing $T$, 
which is also caused by $pp$ and $hh$ configurations 
at low $T$, may slow down the increase of the GDR width. By including 
a simplified $T$-dependent pairing gap in the 
{\scriptsize CASCADE} calculations using the PDM strength functions, 
Ref. \cite{Kaury} 
has improved the agreement between the calculated GDR shapes and
experimental ones.

Very recently the $\gamma$ decays were measured in coincidence with 
$^{17}$O particles scattered inelastically from $^{120}$Sn~\cite{Heckman}.
A GDR width of around 4 MeV has been extracted at $T=$ 1 
MeV, which is smaller than the value of $\sim$ 4.9 
MeV for the GDR width at $T=$ 0. This result and the existing systematic for the 
GDR width in $^{120}$Sn up to $T\simeq$ 2 MeV are significantly lower 
than the prediction by the thermal shape-fluctuation model. 
Based on this, Ref. \cite{Heckman} 
concluded that the narrow width observed in $^{120}$Sn 
at low $T$ is not understood. 
The aim of the present work is to show that it is 
thermal pairing that 
causes the narrow GDR width in $^{120}$Sn at low $T$. 
For this purpose we include the thermal pairing 
gap obtained from the modified-BCS theory~\cite{MHFB,MBCS,MBCS1} in the 
PDM~\cite{PDM1,PDM2,PDM3}, 
and carry out the calculations for the GDR width in $^{120}$Sn at 
$T\leq$ 5 MeV.

The paper is organized as follows.
Section I summarizes the main equations for the GDR including thermal 
pairing within the PDM and discusses 
in detail the physical assumptions of the PDM. 
Section II analyzes the results of calculations of GDR width, 
energy, and cross section for $^{120}$Sn at finite temperature in
comparison with the most recent experimental systematic. The paper
is summarized in the last section, where conclusions are drawn.
\section{Main equations for hot GDR within phonon damping model}
\subsection{Equations for GDR width and energy including thermal pairing}
The quasiparticle representation of the PDM including pairing has 
been already reported in Ref. ~\cite{pygmy}. Therefore we discuss here 
only the final equations, which will be used in numerical calculations.
According to this formalism
the GDR width $\Gamma_{\rm GDR}$ is presented as the sum of quantal 
($\Gamma_{\rm Q}$) and thermal ($\Gamma_{\rm T}$) widths 
as~\cite{PDM1,PDM2,PDM3,pygmy}
\begin{subequations}
\label{width}
\begin{equation}
\Gamma_{\rm GDR}=\Gamma_{\rm Q}+\Gamma_{\rm T}~,
\label{widthtotal}
\end{equation}
\begin{equation}
\Gamma_{\rm Q}=2\pi F_{1}^{2}\sum_{ph}[u_{ph}^{(+)}]^{2}(1-n_{p}-n_{h})
\delta(E_{\rm GDR}-E_{p}-E_{h})~,
\label{GammaQ}
\end{equation}
\begin{equation}
\Gamma_{\rm T}=2\pi F_{2}^{2}\sum_{s>s'}[v_{ss'}^{(-)}]^{2}(n_{s'}-n_{s})
\delta(E_{\rm GDR}-E_{s}+E_{s'})~,
\label{GammaT}
\end{equation}
\end{subequations}
where $(ss')=(pp')$ and $(hh')$ with $p$ and $h$ denoting the orbital 
angular momenta $j_{p}$ and $j_{h}$ for particles and holes, 
respectively. The quantal and thermal widths come from 
the couplings of quasiparticle pairs 
$[\alpha_{p}^{\dagger}\otimes\alpha^{\dagger}_{h}]_{LM}$ and 
$[\alpha_{s}^{\dagger}\otimes{\widetilde{\alpha}}_{s'}]_{LM}$ 
to the GDR, respectively. At zero pairing they correspond to the couplings of 
$ph$ pairs, $[a^{\dagger}_{p}\otimes{\widetilde{a}}_{h}]_{LM}$, and 
$pp$ ($hh$) pairs, 
$[a^{\dagger}_{s}\otimes{\widetilde{a}}_{s'}]_{LM}$,
to the GDR, respectively (The tilde $~{\widetilde{}}$~ denotes 
the time-reversal operation). The quasiparticle energies 
$E_{j}=[(\epsilon_{j}-\bar{\mu})^{2}+\bar{\Delta}^{2}]^{1/2}$ are found
from the modified-BCS equations (39) and (40) of \cite{MBCS1}, which determine
the modified thermal gap $\bar{\Delta}$ and chemical potential 
$\bar{\mu}$ from the single particle energies $\epsilon_{j}$ and 
particle number $N$. From them one defines the Bogoliubov coefficients
$u_{j}$, $v_{j}$, and the combinations 
$u_{ph}^{(+)}=u_{p}v_{h}+v_{p}u_{h}$, and
$v_{ss'}^{(-)}=u_{s}u_{s'}-v_{s}v_{s'}$. 
The quasiparticle occupation number $n_{j}$ is
calculated as~\cite{PDM2}
\begin{equation}
     n_{j}=\frac{1}{\pi}\int_{-\infty}^{\infty}\frac{n_{\rm F}(E)
    \gamma_{j}(E)}{[E-E_{j}-M_{j}(E)]^{2}+\gamma_{j}^{2}(E)}dE~,
    \hspace{6mm} n_{\rm F}(E)=({\rm e}^{E/T}+1)^{-1}~,
     \label{nj}
 \end{equation}
 where $M_{j}(E)$ is 
 the mass operator, and 
 $\gamma_{j}(E)$ is the quasiparticle damping, which is determined as 
 the imaginary part of the complex continuation of $M_{j}(E)$ into the 
 complex energy plane. These quantities appear 
 due to coupling between quasiparticles and the GDR. Their 
 explicit expressions are given by Eqs. (3) and (4) of Ref. 
 \cite{PDM1}, respectively, in which $E_{s}$ is now the quasiparticle 
 energy $E_{j}$.  From Eq. (\ref{nj}) it is seen that
 the functional form for the occupation number  $n_{j}$ 
 is not given by the Fermi-Dirac distribution $n_{\rm F}(E_{j})$ 
 for non-interacting 
 quasiparticles. It can be approximately to be so
 if the quasiparticle damping $\gamma_{j}(E)$ is sufficiently small 
 so that the Breit-Wigner-like kernel under the integration can be replaced with 
 the $\delta$-function. Equation (\ref{nj}) also implies a zero value 
 for $n_{j}$ in the ground state, i.e. $n_{j}(T=0)=$ 0. 
 In general, it is not the case because of ground-state 
 correlations (See. e.g Refs. \cite{MBCS,RRPA,RRPA1}). They lead to 
 $n_{j}(T=0)\neq$ 0, which should be found by solving 
 self-consistently a set of nonlinear equations within the renormalized random-phase 
 approximation (renormalized RPA). Within the RPA the equation for 
 $n_{j}$ yields the approximate expression 
 $n_{j}(T=0)\simeq\sum_{Jij'}(2J+1)/(2j+1)[Y_{jj'}^{(Ji)}]^{2}$, 
 where $Y_{jj'}^{(Ji)}$ is the RPA backward-going amplitude ($J$ is 
 the multipolarity). 
 Since for collective high-lying excitations such as GDR one has 
 $|Y_{jj'}^{(Ji)}|\ll$ 1, we expect the value $n_{j}(T=0)$ to be negligible. 
  
The 
GDR energy $E_{\rm GDR}$ is found as the solution of the equation
\begin{equation}
\omega-\omega_{q}-P(\omega)= 0,
\label{EGDR}
\end{equation}
where $\omega_{q}$ is the unperturbed phonon energy, and 
$P(\omega)$ is the polarization operator:
\begin{equation}
 P(E)=F_{1}^{2}\sum_{ph}\frac{[u_{ph}^{(+)}]^{2}
 (1-n_{p}-n_{h})}
 {E-E_{p}-E_{h}}-F_{2}^{2}\sum_{s>s'}
 \frac{[v_{ss'}^{(-)}]^{2}(n_{s}-n_{s'})}
 {E-E_{s}+E_{s'}}~,
 \label{P}
 \end{equation} 
Note that, in general, there are also backward-going 
processes leading to the terms $\sim \delta(\omega+E_{p}+E_{h})$ and 
$\delta(\omega+E_{s}-E_{s'})$ as has been shown in Eqs. (14) and 
(15) of Ref. \cite{pygmy}. However, as 
the maximum of these terms is located at negative energy
$\omega=-(E_{p}+E_{h})<$ 0, and $\omega=-(E_{s}-E_{s'})<$ 0, 
respectively, their 
contribution  
to the GDR, which is located at $\omega=E_{\rm GDR}\gg$ 1 MeV, is negligible.
Therefore these backward-going processes are omitted here.
It is now easy to see that, at zero pairing $\bar{\Delta}=$ 0, one has
$u_{p}=$ 1, $v_{p}=$ 0, $u_{h}=$ 0, $v_{h}=$ 1 so that
$[u^{(+)}_{ph}]^{2}=$ 1, $[v^{(-)}_{ss'}]^{2}=$ 1. As for the 
single-particle occupation 
number $f_{j}$, one obtains $f_{h}=1-n_{h}$ and $f_{p}=n_{p}$. 
The PDM equations for $\bar{\Delta}=$ 0
in Ref. \cite{PDM1,PDM2} are then easily recovered from Eqs. (\ref{width}) -- (\ref{P}). 
\subsection{Assumptions of phonon damping model}
The PDM is based on the following assumptions:

i) The matrix elements for the coupling of GDR to non-collective $ph$ 
configurations, which causes the quantal width $\Gamma_{\rm Q}$ 
(\ref{GammaQ}), are all equal to $F_{1}$. 
Those for the coupling of GDR to $pp$ ($hh$), which causes the thermal 
width $\Gamma_{\rm T}$ (\ref{GammaT}), 
are all equal to $F_{2}$.
The assumption of a constant coupling strength
is well justified when the width of a collective mode is much 
smaller than the energy range $\Delta E$ (of order of $E_{\rm GDR}$) 
over which this mode is coupled to the background states (the 
so-called weak coupling limit discussed in Refs. \cite{Ze,Lauritzen}).

ii) It is well established that 
the microscopic mechanism of the quantal (spreading) width 
$\Gamma_{\rm Q}$ (\ref{GammaQ}) comes from 
quantal coupling of $ph$ configurations to more complicated ones, 
such as $2p2h$ ones. The calculations performed 
in Refs. \cite{Bortignon,DangGDR} 
within two independent microscopic models, where such couplings to 
$2p2h$ configurations were explicitly included, have shown that $\Gamma_{\rm Q}$ 
depends weakly on $T$. The microscopic 
study in Ref. \cite{Lauritzen}, where 
an hierarchy of states of increasing complexity located around $E_{\rm 
GDR}$ is considered, has also confirmed the nearly 
constancy of $\Gamma_{\rm Q}$. It also indicated that the
width of a collective vibration does not depend on the detailed 
coupling to the compound nuclear eigenstates. Therefore, 
in order to avoid complicate numerical calculations, which are not 
essential for the increase of $\Gamma_{\rm GDR}$ 
at $T\neq$ 0, such microscopic mechanism is not 
included within PDM, assuming
that $\Gamma_{\rm Q}$ at $T=$ 0 is known. The model parameters
are then chosen so that the calculated $\Gamma_{\rm Q}$ and $E_{\rm 
GDR}$ reproduce the corresponding experimental values 
at $T=$ 0 (See below). 

Assumption (i) is satisfied for $F_{1}$ since the quantal width 
$\Gamma_{\rm Q}$ does not exceed 4.9 MeV. The PDM calculations 
in fact have shown that $\Gamma_{\rm Q}$ 
decreases from 4.9 MeV at $T=$ 0 to around 2.5  
MeV at $T=$ 5 MeV for $^{120}$Sn due to thermal effects in 
the factor $1-n_{p}-n_{h}$  
(See the dashed line in Fig. 1 (a) of \cite{PDM1} for zero pairing).
A similar trend was also observed in the microscopic calculations of 
Ref. \cite{Bortignon}, where it was found that
the GDR width at $T=$ 3 MeV is in fact 
smaller that at $T=$ 0 (See Figs. 9 and 10 of Ref. \cite{Bortignon} 
an the discussion therein). 
That's why $\Gamma_{\rm Q}(T=0)$ cannot be 
simply taken as a parameter uniformly added to what is
calculated for $\Gamma_{\rm T}$ at $T\neq$ 0 since
a width $\Gamma_{\rm Q}(T)=\Gamma_{\rm Q}(T=0)$ would lead to a 
larger value for the total 
width $\Gamma_{\rm GDR}$ (\ref{widthtotal}) at higher $T$, worsening 
the agreement with the data.  

Assumption (i) becomes poor for $F_{2}$ at $T\agt$ 3 MeV, when 
the thermal width $\Gamma_{\rm T}$ is larger than 10 MeV 
(See the dotted line in Fig. 
1 (a) of Ref. \cite{PDM1}). Within
such a large width one expects a considerable change of the 
level density of background states. 
To be quantitatively precise, one needs to use 
a self-consistent theory for the strength function, which includes 
the coupling to doorway states as in Ref. \cite{Lauritzen}. Such 
theory is valid for any ratio of $\Gamma_{\rm GDR}/\Delta E$.
However, from assumption (ii) it also follows that 
the increase of $\Gamma_{\rm GDR}$ is 
now driven mainly by the thermal width 
$\Gamma_{\rm T}$ due to the 
factor $n_{s'}-n_{s}$.
The change of $n_{j}$ implies a change of the quasiparticle 
entropy $S_{\rm qp}$. The latter is ultimately related to the
change of the level density of background states within the 
realistic mean-field basis.
This can be seen as follows.
The complexity of the background states is measured by 
the information entropy 
$S_{\rm inf}$ of individual wave functions, which reflects 
the complicated relationship between the eigenbasis and 
representation basis. Meanwhile, the thermodynamic entropy $S_{\rm th}$ 
of the total system is directly determined by its statistical weight
$\Omega(E)=\rho(E)\delta(E)$ as $S_{\rm th}=\ln\Omega(E)$, where 
$\rho(E)$ is the level density. In a situation with incomplete 
information, such as in the statistical description of hot nuclei 
considered here, individual 
compound systems are replaced with a 
grand canonical ensemble of nuclei in thermal equilibrium.
The probability for a quantum system to have a given eigenenergy 
is determined by the density matrix ${\cal D}$ 
rather than by a pure wave function. 
The expectation value $\langle{\cal O}\rangle$ of an observable 
${\cal O}$ is given as the statistical average over the grand 
canonical ensemble $\langle{\cal O}\rangle={\rm Tr}({\cal D}{\cal 
O})$, which is derived from the maximum of 
the thermodynamic entropy $S_{\rm th}=-{\rm Tr}({\cal D}\ln{\cal D})$.
The modern shell-model calculations in Ref. \cite{Ze} have shown that
these three apparently different entropies, $S_{\rm qp}$, $S_{\rm 
inf}$, and $S_{\rm th}$, behave very similarly for 
the majority of states in the realistic mean field consistent with 
residual interactions (See column II of Fig. 56 in Ref. \cite{Ze} and 
the discussion therein).  Significant differences between them take 
place only when the residual interaction beyond the mean field 
is very weak (See column I of Fig. 56 in \cite{Ze}) or
when the quasiparticle mean-field is absent ($S_{\rm inf}$ reaches 
its chaotic limit)
(See column III of Fig. 56 in \cite{Ze}).
This might be the reason why the numerical results 
performed so far within the zero-pairing PDM~\cite{PDM1,PDM2,PDM3} 
using assumptions (i) and (ii) 
fit fairy well the 
experimental systematic for the GDR width 
including the existing data 
on the width saturation at $T\agt$ 3 MeV. 
This is partly also due to the 
large experimental errorbars for the extracted GDR width 
at high $T$ (See, e.g. Refs. ~\cite{Bracco,Gaar}). 
Therefore, at $T\agt$ 3 
MeV the results obtained under these assumptions
should be considered as qualitative. This does not affect our present study 
of the pairing effect as the latter is significant only at low 
temperature ($T<$ 2.5 MeV).

Within assumptions (i) and (ii) the model has only three 
$T$-independent parameters, which are the unperturbed phonon energy 
$\omega_{q}$, $F_{1}$, and  $F_{2}$. 
The parameters $\omega_{q}$ and $F_{1}$ are 
chosen so that after the $ph$-GDR coupling is switched on,
the calculated GDR energy $E_{\rm GDR}$ and width $\Gamma_{\rm GDR}$ 
reproduce the corresponding experimental values for GDR on 
the ground-state. At $T\neq$ 0, the coupling to $pp$ and $hh$ 
configurations is activated.
The $F_{2}$ parameter is then fixed at $T=$ 0 
so that the GDR energy $E_{\rm GDR}$ 
does not change appreciably with varying $T$. 
The values of the PDM parameters for $^{120}$Sn 
are given in \cite{PDM2} for the zero-pairing case. 

In Ref. \cite{PDM2} we have presented 
an argument that, in our opinion, the effect due to 
coupling of GDR to noncollective $ph$, $pp$ and $hh$ 
configurations at $T\neq$ 0 within the PDM is 
tantamount to that of thermal shape fluctuations.
This has been demonstrated by expanding 
the coupling of GDR phonon to noncollective $ph$, $pp$, 
and $hh$ configurations into couplings to different multipole fields
(See pages 437 and 438 of Ref. \cite{PDM2}).
In this expansion the $pp$ ($hh$)-pair operator
$B_{ss'}^{\dagger}\equiv a_{s}^{\dagger}a_{s'}$ is expanded in terms
of the tensor products of two $ph$-pair operators (See mappings (22) and 
(23) of Ref. \cite{RRPA1}). 
Each $ph$-pair operator $B_{ph}^{\dagger}\equiv a_{p}^{\dagger}a_{h}$ 
can be then expressed in terms 
of the RPA phonon operators $Q_{q}^{\dagger}$ 
and $Q_{q}$ with RPA amplitudes 
$X_{ph}^{q}$ and $Y_{ph}^{q}$. This leads to Eq. (2.43) in Ref. \cite{PDM2} 
for the part of the PDM
Hamiltonian, which describes coupling between the phonon and 
single-particle fields. Hence if
$Q_{q}^{\dagger}$ and $Q_{q}$ are GDR phonon operators, 
$\{Q_{q_{1}}^{\dagger},Q_{q_{1}}\}$ and $\{Q_{q_{2}}^{\dagger},Q_{q_{2}}\}$ 
in Eq. (2.43) of Ref. \cite{PDM2} can have the moment and parity 
equal to $(1^{-}, 2^{+})$, $(2^{+}, 3^{-})$, etc. to preserve the total 
momentum $\lambda^{\pi}=1^{-}$. Therefore, 
although $F_{ss'}^{(q)}$ are dipole matrix elements, the 
amplitudes $X_{ph}^{q_{i}}$ and $Y_{ph}^{q_{i}}$ ($i=1,2$) can be calculated 
microscopically, using the 
dipole-dipole, quadrupole-quadrupole, octupole-octupole, etc. 
components of residual interaction.  This means that coupling to $pp$ and 
$hh$ configurations already includes in principle the coupling to different 
multipole-multipole fields via multiphonon configuration mixing at 
$T\neq$ 0 \footnote{In Ref. \cite{PDM4} a version of PDM, which explicitly 
includes coupling to two-phonon configurations in the second 
order of the interaction vertex, was proposed. 
The expressions (2.17) and (2.18) of Ref. \cite{PDM4} derived  
for the polarization operator show that the
width increase is still driven mainly by the factor $(n_{s}-n_{s'})$.
The calculations with the GDR coupled to the 
first quadrupole phonon $2^{+}_{1}$ required
the energy $\omega_{2^{+}_{1}}$ and the ratio 
$r=F^{(2)}_{i}/F^{(1)}_{i}$
$(i=1,2)$ to be fixed as additional parameters.
A similar quality for the description of the experimental data for the 
temperature dependence of the 
GDR width has been restored after a reduction of the dipole matrix elements
$F_{1}^{(1)}\equiv F_{1}$ and $F_{2}^{(1)}\equiv F_{2}$.}.

{\scriptsize CASCADE} calculations 
using PDM strength functions  
have produced the GDR shapes in good
agreement with the experimental 
data for $^{120}$Sn at $T>$ 1.54 MeV (See Fig. 2 of Ref. \cite{Kaury}).
The discrepancy at $T\leq$ 1.54 MeV is due to omission or improperly 
inclusion of pairing
at low $T$, which is now studied in the present work.
The splitting of GDR into two peaks 
is also clearly seen in the PDM calculations for $^{106}$Sn~\cite{Sn106}. 
These evidences are in favor of the
argument discussed above and in Ref. \cite{PDM2}.

Nonetheless, we recognize that the issue of whether coupling to $pp$ and $hh$ 
configurations at $T\neq$ 0 is a microscopic (although indirect) 
interpretation of thermal shape fluctuations within the PDM is still 
not settled. The question on whether thermal shape fluctuations need 
to be included additionally within PDM or not remains to be investigated.
The thermal shape-fluctuation model
calculates the time-correlation function of the GDR 
by replacing
the micro-canonical ensemble with an ensemble of macroscopic
variables, which are the deformation parameters $(\beta,\gamma)$ 
in the body-fixed (principal axes) frame of reference 
(intrinsic frame). It
parametrizes a priori the dipole 
correlation tensor
by a frequency 
$E_{k}=E_{0}\exp[-\sqrt{5/4\pi}\beta\cos(\gamma+\frac{2}{3}\pi k)]$ and 
a width equal to 
$\Gamma_{k}=\Gamma_{0}(E_{k}/E_{0})^{1.8}$ along each $i$th semiaxis
~\cite{TFM}.  The effect of 
thermal fluctuations in this model is included via fluctuations of shapes by 
employing the macroscopic Landau theory of phase 
transitions~\cite{Landau}.
Therefore, an explicit inclusion of thermal shape 
fluctuations in the PDM will bring in additional degrees of freedom, 
which increase the number of parameters of the model. 
At the same time, 
in a way similar to that mentioned in the footnote 
above, this 
will certainly require a renormalization of the existing 
parameters of the PDM to restore the agreement with the 
experimental systematic. 
While this issue is left open for future 
study, it does not affect the study of the role 
of thermal pairing in the present paper, 
since, as will be seen below, in order to describe the GDR 
width at low $T$,
thermal pairing is necessary to be included in any model, 
whether it is the PDM or 
thermal shape-fluctuation one.
\section{Analysis of numerical calculations}
\subsection{Role of thermal pairing gap at 
low temperature}

Shown in Fig. \ref{gap} is the $T$ dependence of the neutron 
pairing gap $\bar{\Delta}_{\nu}$ for $^{120}$Sn, which is 
obtained from the modified-BCS equations~\cite{MHFB,MBCS,MBCS1} using the 
single-particle energies 
determined within the Woods-Saxon potential at $T=$ 0. They span a  
space from $\sim$ -40 MeV up to $\sim$ 17 MeV including 7 major 
shells and $1j_{15/2}$, $1i_{11/2}$, and $1k_{17/2}$ levels. 
The pairing parameter $G_{\nu}$ is chosen to be equal to 0.13 MeV, which 
yields $\bar{\Delta}(T=0)\equiv\bar{\Delta}(0)\simeq$ 1.4 MeV. 
In difference with the BCS gap (dotted line), 
which collapses at $T_{\rm c}\simeq$ 
0.79 MeV, the 
gap $\bar{\Delta}$ (solid line) does not vanish, but
decreases monotonously with increasing $T$ at $T\agt$ 1 MeV 
resulting in a long tail up to $T\simeq$ 5 MeV.
This behavior is caused by the thermal fluctuation 
of quasiparticle number,
$\delta{\cal N}^{2}\equiv\sum_{j}\delta{\cal N}_{j}^{2}$, where
$\delta{\cal N}_{j}^{2}=n_{j}(1-n_{j})$ is 
the quasiparticle-number fluctuation 
on $j$-th orbital. The latter is incorporated as $\delta{\cal N}_{j}$ 
in the modified-BCS gap $\bar{\Delta}$ (See the 
last term at the r.h.s of Eq. (39) of Ref. \cite{MBCS1}). 

To analyze the qualitative effect of thermal pairing on the GDR width 
we plot
the usual single-particle occupation number 
$f_{j}=u_{j}^{2}n_{j}+v_{j}^{2}(1-n_{j})$ with 
$n_{j}$ obtained within the modified-BCS theory 
($\gamma_{j}(E)=$ 0) as a function 
of single-particle energy $\epsilon_{j}$ for the neutron levels around 
the chemical potential in Fig. \ref{fn}. It is seen that, in general, 
the pairing effect always goes counter the 
temperature effect on $f_{j}$, causing a 
steeper dependence of $f_{j}$ on $\epsilon_{j}$. 
Decreasing with increasing $T$, this difference becomes 
small at $T\geq$ 3 MeV. 
Since a smoother $f_{j}$ enhances the $pp$ 
and $hh$ transitions leading to the thermal width 
$\Gamma_{\rm T}$, 
pairing should reduce the GDR width, and this reduction 
is expected to be stronger at a lower $T$, provided the GDR 
energy $E_{\rm GDR}$ is the same. 
A deviation from this general rule 
is seen at very low $T\simeq$ 0.1 MeV, where the temperature effect is 
still so weak that $f_{j}$ 
obtained at $\bar{\Delta}\neq$ 0 (solid line) 
is smoother than that obtained at zero pairing (dotted line). 

To get an insight into the detail of the change of GDR width at low 
$T$ we show in Fig. \ref{uvn}
the combinations $[u_{ph}^{(+)}]^{2}$ and $[v_{hh'}^{(-)}]^{2}$
of the Bogoliubov coefficients $u_{p}$, $v_{h}$, and $v_{h'}$ 
together with the factors $(1-n_{p}-n_{h})$ 
and $(n_{h}-n_{h'})$ as well as their zero-pairing counterparts,
$f_{h}-f_{p}$ and $f_{h'}-f_{h}$ for the particle $p=2j_{7/2}$, 
hole $h=2d_{3/2}$,
and $h'=1d_{5/2}$ orbits as functions of $T$. 
The hole orbit $h=2d_{3/2}$ is located just 
below the chemical potential. Therefore the pairing effect is strongest 
for the $ph$ and $hh$ configurations involving this orbit. 
This figure shows a sharp increase of $[u_{ph}^{(+)}]^{2}$ and 
$[v_{ss'}^{(+)}]^{2}$ at $T\alt$ 1 -- 2 MeV due 
to a steep slope of the pairing gap, showing a strong pairing effect. 
At very low $T$, the numerator 
of the polarization operator (\ref{P}) is close to
$[u_{ph}^{(+)}]^{2}(E_{p}+E_{h})$ because $1-n_{p}-n_{h}\simeq$ 1, 
while the thermal part $\sim (n_{s}-n_{s'})\simeq$ 0. This value is 
equal to 3.67 MeV at $T=$ 0.1 MeV, which is smaller than 
$(\epsilon_{h}-\epsilon_{p})=$ 4.63 MeV obtained at $\bar{\Delta}=$ 0.
The denominator of (\ref{P}) 
is also smaller than that obtained 
at $\bar{\Delta}=$ 0
because $E_{p}+E_{h}>\epsilon_{h}-\epsilon_{p}$ due to the gap.
Therefore, at very low $T$, 
pairing may lead even to a smaller GDR energy. On the other hand, as
$[u^{(+)}_{ph}]^{(2)}(1-n_{p}-n_{h})$ and 
$[v^{(-)}_{ss'}]^{2}(n_{s}-n_{s'})$ are also smaller than $f_{h}-f_{p}$ 
and $f_{s'}-f_{s}$, respectively, the competition of these effects
in Eq. (\ref{width}) 
can result in a larger width in the very low-$T$ region. 
As $T$ increases, the factor $1-n_{p}-n_{h}$ decreases while 
$[u_{ph}^{(+)}]^{2}$ increases to reach 1 at $T\agt$ 2 MeV 
because of the decreasing gap. This leads to the 
decrease of the quantal width $\Gamma_{\rm Q}$. 
At the same time, coupling to $pp$ and $hh$ 
configurations starts to contribute due to the factor $(n_{s}-n_{s'})$. 
The combination $[v_{ss'}^{(-)}]^{2}$ 
also increases with $T$ and reach 1 at $T\agt$ 3
MeV. As the result, thermal width 
$\Gamma_{\rm T}$ starts to give an increasing contribution with $T$.
However, as compared to their 
zero-pairing counterparts, $f_{h}-f_{p}$ and $f_{s'}-f_{s}$,
the decrease of the quantal part $\sim (1-n_{p}-n_{h})$ and increase
of the thermal part $\sim (n_{s}-n_{s'})$ are much more moderate
with increasing $T$ up to 1 MeV. On the contrary, at 1 $\alt T\alt$ 
3 MeV the decrease $(1-n_{p}-n_{h})$ and increase of
$(n_{s}-n_{s'})$ with increasing $T$ are steeper than their 
counterparts at $\bar{\Delta}=$ 0.
At $T>$ 3 -- 4 MeV the total width approaches the saturation 
because of the dominating contribution of $\Gamma_{\rm T}$, which 
ceases to increase due to the $T$ dependence of $n_{s}-n_{s'}$ 
shown in Fig. \ref{uvn} (b)~\cite{PDM1,PDM2}.
\subsection{Temperature dependence of GDR width and energy}
The GDR width $\Gamma_{\rm GDR}$ and energy $E_{\rm GDR}$ for $^{120}$Sn 
were calculated from 
Eqs. (\ref{width}) and (\ref{EGDR}), respectively, 
using the same set of PDM parameters 
$\omega_{q}$, $F_{1}$, and $F_{2}$, which have been chosen  
for the zero-pairing case~\cite{PDM1,PDM2} (set A). The effect of 
quasiparticle damping is included in the 
calculations by using Eq. (\ref{nj}). The results are shown
as the thin solid lines in Fig. \ref{widthSn}.
As seen from Fig. \ref{widthSn} (b), 
the oscillation of GDR energy $E_{\rm GDR}$ with varying $T$ 
occurs within the range of $\sim\pm 1.5$ MeV, which is 
wider compared with that obtained neglecting pairing. The latter is 
almost independent on $T$ (dashed line in Fig. \ref{widthSn} (b)). 
As expected from the 
discussion above, the GDR energy 
$E_{\rm GDR}$ at 
$T=$ 0.1 MeV drops to 14 MeV, i.e. by 1.4 MeV lower than 
the GDR energy measured on the ground state.
The GDR width increases to 5.3 MeV compared to 4.9 MeV on the ground 
state as shown in Fig. \ref{widthSn} (a). 
At 0 $\alt T\alt$ 0.5 MeV, 
the above-mentioned competition between the decreasing quantal and 
increasing thermal widths makes the total width decrease first to reach a 
minimum of 3.4 MeV at $T\simeq$ 0.2 MeV then increase again with $T$.
At $T \agt T_{\rm c}$ the width only increases with $T$. At 
1 $\alt T\alt$ 3 MeV the GDR width obtained including pairing 
is smaller than the one obtained neglecting pairing (dashed line in Fig. 
\ref{widthSn} (a)), but this
difference decreases with increasing $T$ so that at $T>$ 3 MeV, 
when the gap $\bar{\Delta}$ becomes small, both values nearly coincide. This 
improves significantly the agreement with the experimental systematic 
at 1 $\alt T\alt$ 2.5 MeV.
In order to have the same value of 4.9 MeV for the GDR width at 
$T\simeq$ 0, we 
also carried out the calculation using slightly 
readjusted values $F_{1}'=0.96 F_{1}$ and $F_{2}'=1.03 F_{2}$ while
keeping the same $\omega_{q}$ (set B). The 
result obtained is shown in the same Fig. \ref{widthSn} as the thick 
solid lines. The GDR energy $E_{\rm GDR}$ moves up to 16.6 MeV at 
$T=$ 0.1 MeV and at $T_{\rm c}\alt T\alt$ 1.2 MeV 
in agreement with the value of 16.5 $\pm$ 0.7 MeV extracted at $T=$ 1 
MeV in Ref. \cite{Heckman}. 
The width at $T=$ 1 MeV also becomes 
slightly smaller, which agrees quite well with the latest 
experimental point~\cite{Heckman}. At $T>$ 1 MeV the results obtained 
using two parameter sets, A and B, 
are nearly the same. The effect of quantal 
fluctuations $\delta N^{2}$ of particle number within the BCS theory at 
$T=$ 0, however, is neglected in these results.
To be precise, this effect
should be included using the particle-number 
projection method at finite $T$. However, the latter is so computationally 
intensive that the calculations were carried out so far only within
schematic models (See, e.g. \cite{Rossignoli}), or one 
major shell for nuclei with A $\leq$ 
60 as in the Shell-Model Monte-Carlo method~\cite{Alhassid}. 
Therefore, for the limited purpose of the present study, 
assuming that $\delta N^{2}\gg$ 1, we applied the approximated 
projection at $T=$ 0 proposed in Ref. \cite{Mikhailov}, 
which leads to the renormalization of the gap as
$\widetilde{\Delta}(T)=[1+1/\delta N^{2}]\bar{\Delta}$ with
$\delta N^{2}=\bar{\Delta}(0)^{2}\sum_{j}(j+1/2)/
[(\epsilon_{j}-\bar{\mu})^{2}+\bar{\Delta}(0)^{2}]$~\cite{DangZ}. This yields
$\widetilde\Delta(T=0)\simeq$ 1.5 MeV ($\delta N\simeq\pm$ 4). 
The PDM results obtained using the 
gap $\widetilde{\Delta}$ and the parameter set B 
are shown in the same 
Fig. \ref{widthSn} as the thick dotted lines. The GDR width becomes 5 MeV 
with $E_{\rm GDR}=$ 15.3 MeV at $T=$ 0 in good agreement with the GDR 
parameters extracted on the ground state. 
It is seen that the fluctuation of the 
width at $T\alt$ 0.5 MeV is largely suppressed by using this 
renormalized gap $\widetilde{\Delta}$. For comparison, the 
predictions by two versions of thermal shape-fluctuation 
model~\cite{TFM,Kuznesov} are
also plotted in Fig. \ref{widthSn} (a) as the dash-dotted~\cite{TFM}
and thin dotted~\cite{Kuznesov} lines.
It is seen that these predictions, in particular the one given by the 
phenomenological version in Ref. \cite{Kuznesov}, 
significantly overestimate the GDR width at low 
temperature $T\alt$ 1.3 MeV. The predicted overall increase of the width 
is not as steep as the experimental systematic and the PDM prediction.
The curvature of the trend is also 
opposite to the experimental one and that 
given by the PDM.
It is, therefore,
highly desirable to see how the prediction by the thermal shape-fluctuation 
model would change by taking into account 
the effect of thermal pairing gap discussed in the present work in 
combination with the use of a specific Hamiltonian to calculate every 
quantity~\cite{Ansari}.   
\subsection{Effect of thermal pairing on GDR cross section at low 
temperature}
Shown in Fig. \ref{stre} are GDR cross-sections obtained 
for $^{120}$Sn using Eq. (1) of Ref. 
\cite{Kaury}.
The experimental cross-section are taken from
Fig. 2 of Ref. \cite{Kaury}. They have been 
generated by {\scriptsize{CASCADE}}
at excitation energy $E^{*}=$ 30 and 50 MeV, which correspond to
$T_{\rm max}=$ 1.24, and 1.54 MeV, respectively.
The theoretical cross-sections have been obtained using the
PDM strength function $S_{\rm GDR}(E_{\gamma})$ 
from Eq. (2.22) of Ref. \cite{PDM3} at $T=T_{\rm max}$.
This is the low temperatures region, at which discrepancies
are most pronounced 
between theory and experiment.
(A divided spectrum free from detector response at 
$T=$ 1 MeV is not available in Ref. \cite{Heckman}). 
From this figure it is seen that 
thermal pairing clearly offers a 
better fit to the experimental line shape of the GDR at low temperature.
As has been discussed in Ref. \cite{Kaury}, 
for an absolute comparison, the PDM strength functions for all decay 
steps starting from $T_{\rm max}$ down to $T\simeq$ 0 MeV should be 
included in the {\scriptsize{CASCADE}} to generate a divided spectra, 
which can be directly compared with the experimental ones.
It is our wish that such calculations can be carried out in 
collarboration with the authors of \cite{Heckman} in the near future.
\section{Conclusions}
In this paper we have included the pairing gap, determined 
within the modified-BCS theory~\cite{MBCS,MBCS1}, in the 
PDM to calculate the width
of GDR in $^{120}$Sn at $T\leq$ 5 MeV. 
In difference with the 
gap given within the conventional BCS theory, 
which collapses at $T_{\rm c}\simeq$ 0.79 MeV, 
the modified-BCS gap never vanishes, but monotonously decreases with 
increasing $T$ up to $T\simeq$ 5 MeV. The results obtained show that
thermal pairing indeed plays an important role in lowering the width
at $T\alt$ 2 MeV as compared to the value obtained without pairing. 
This improves significantly the 
overall agreement between theory and experiment, 
including the width at $T=$ 1 MeV 
extracted in the latest experiment~\cite{Heckman}.
\acknowledgements
The numerical calculations were carried out using the FORTRAN IMSL 
Library 3.0 by Visual Numerics on the Alpha server 800 5/500 at the 
Division of Computer and Information of RIKEN.
%\newpage %Just because of unusual number of tables stacked at end

\newpage
%{\bf Figure captions}
%%%%%%%%%%%%%%%%%%%%%%% Figure 1 %%%%%%%%%%%%%%%%%%%%%%%%%%%%%%
\begin{figure}
fig1.gif
    \caption{\label{gap}Neutron pairing gap 
as a function of $T$.
Solid and dotted lines show the modified-BCS gap $\bar{\Delta}$ and BCS 
gap, respectively.}
\end{figure}
%%%%%%%%%%%%%%%%%%%%%%%%%%%%%%%%%%%%%%%%%%%%%%%%%%%%%%%%%%%%%%%%
%%%%%%%%%%%%%%%%%%%%%%% Figure 2 %%%%%%%%%%%%%%%%%%%%%%%%%%%%%%
\begin{figure} 
fig2.gif
\caption{\label{fn}Single-particle occupation number $f_{j}$ as  
a function of $\epsilon_{j}$ for the neutron 
levels around the chemical potential at $T=$ 0.1, 1, and 3 MeV. 
Results obtained including and without pairing are shown by solid and 
dotted lines, respectively (A thicker line corresponds to a higher $T$). 
The horizontal 
dashed line at $\sim$ -6 MeV shows the chemical potential at 
$\bar{\Delta}=$ 0 and $T=$ 0.}
\end{figure}
%%%%%%%%%%%%%%%%%%%%%%%%%%%%%%%%%%%%%%%%%%%%%%%%%%%%%%%%%%%%%%%%                                                                                                                                                                                                                                                                           
%%%%%%%%%%%%%%%%%%%%%%% Figure 3 %%%%%%%%%%%%%%%%%%%%%%%%%%%%%%
\begin{figure}                                                             
fig3.gif
\caption{\label{uvn}Combinations of Bogoliubov coefficients (a) and
quasiparticle occupations numbers (b). 
In (a): The thin and thick lines show $[u_{ph}^{(+)}]^{2}$ 
and $[v_{hh'}^{(-)}]^{2}$, respectively, for the orbits 
$p=2j_{7/2}$, $h=2d_{3/2}$, and $h'=1d_{5/2}$.
In (b); The corresponding factors $1-n_{p}-n_{h}$ (thin line) 
and $n_{h}-n_{h'}$ (thick line) are shown. The factors $f_{h}-f_{p}$ 
and $f_{h'}-f_{h}$ for the zero-pairing case are shown as 
the dotted and dashed lines, respectively.}
\end{figure}
%%%%%%%%%%%%%%%%%%%%%%%%%%%%%%%%%%%%%%%%%%%%%%%%%%%%%%%%%%%%%%%%  
%%%%%%%%%%%%%%%%%%%%%%% Figure 4 %%%%%%%%%%%%%%%%%%%%%%%%%%%%%%
\begin{figure}                                                             
fig4.gif  
\caption{\label{widthSn}GDR width $\Gamma_{\rm GDR}$ (a) and energy 
$E_{\rm GDR}$ (b) as functions of 
$T$ for $^{120}$Sn. The black solid lines show the 
PDM results obtained neglecting pairing (In (a) it is the same as 
the solid line with diamonds 
from Fig. 1 (a) of Ref. \cite{PDM1}). 
The green and red lines are the PDM results 
including the gap $\bar{\Delta}$, which are 
obtained using the parameter sets A and B, respectively.
The blue lines are the PDM results including the renormalized gap 
$\widetilde{\Delta}$ (See text).
Solid circles are the low-$T$ data from 
Ref. \cite{Heckman}. Crosses and open triangles in (a) are from Fig. 4 of 
Ref. \cite{Kuznesov}. The corresponding GDR energies decrease 
from 16 MeV to 14.5 MeV with 
increasing $T$ as shown in the shaded 
rectangle in (b). Solid upside-down triangles are data from 
Ref. \cite{Kelly}. Open squares and stars are high-$T$ data for $^{110}$Sn 
from Refs. \cite{Bracco} and 
\cite{Gaar}, respectively. Data at $T=$ 0 are for 
GDR built on the ground state of tin isotopes with 
masses $A=$ 116 -- 124 from Ref. \cite{Berman}. The predictions by two 
versions of the thermal shape-fluctuation model are shown in (a) as the 
dash-dotted~\cite{TFM} 
and thin dotted~\cite{Kuznesov} lines, respectively.}
\end{figure}
%%%%%%%%%%%%%%%%%%%%%%%%%%%%%%%%%%%%%%%%%%%%%%%%%%%%%%%%%%%%%%%%                                                                
%%%%%%%%%%%%%%%%%%%%%%% Figure 5 %%%%%%%%%%%%%%%%%%%%%%%%%%%%%%
\begin{figure}                                                             
fig5.gif
    \caption{\label{streSn} Experimental (shaded areas) 
and theoretical divided spectra 
obtained without pairing (black solid lines) and 
including the gap $\bar{\Delta}$ (red lines) as for
the red lines in Fig. \ref{widthSn}.
\label{stre}}
\end{figure}
%%%%%%%%%%%%%%%%%%%%%%%%%%%%%%%%%%%%%%%%%%%%%%%%%%%%%%%%%%%%%%%%                                                                  
\end{document}